\documentclass[twocolumn,showpacs,preprintnumbers,amsmath,amssymb]{revtex4}

\usepackage{graphicx}
\usepackage{dcolumn}

\newcommand{\Hfill}{{H_\text{fill}}}

\begin{document}


\title{Shear zones in granular media: 3D Contact Dynamics simulation}

\author{Alexander Ries}
 \affiliation{Department of Physics, University Duisburg-Essen,
 D-47048 Duisburg, Germany}

\author{Tam\'as Unger}
\affiliation{%
Department of Theoretical Physics, 
Budapest University of Technology and Economics,
H-1111 Budapest, Hungary}

\author{Dietrich E. Wolf}
 \affiliation{Department of Physics, University Duisburg-Essen,
 D-47048 Duisburg, Germany}

\date{\today}

\begin{abstract}

Shear zone formation is investigated in slow 3D shear flows. We simulate
the linear version of the split-bottom shear cell. It is shown that the
same type of wide shear zones is achieved in the presence as well as
in the absence of gravity. We investigate the relaxation of the material
towards a stationary flow and analyze the stress and the velocity fields.
We provide the functional form of the widening of the shear zone inside the
bulk. We discuss the growth of the region where the material is in critical
state. It is found that the growth of the critical zone is responsible for
the initial transient of the shear zone.
\end{abstract}

\pacs{47.57.Gc, 45.70.-n, 83.50.Ax}

\maketitle

\section{Introduction}

Granular materials consist of macroscopic grains which have simple
interactions and obey the laws of classical mechanics.  The resulting
behavior, however, can be surprisingly complex. This is shown by a large
variety of phenomena \cite{Aranson06,deGennes99} which, in most cases,
present great challenges to theoretical descriptions.

%
%
%

In the present paper we focus on one of the unsettled problems of granular
media: the quasi-static rheology \cite{GDRMiDi04}. The flow is called
quasi-static when inertia effects are negligible. This can be achieved by a
combination of large pressure and low deformation rate.

An important property observed for this type of flows is that stresses
become independent of the deformation rate. This is remarkable because it
is the rate dependence that is naturally expected to be responsible for the
rheology. For quasi-static flows the link between stress and deformation
rate is missing and we lack the constitutive law which could describe the
deformation field in the material.

An experimental setup which is particularly suited to provide insight
into quasi-static granular flow is the split-bottom shear cell. It has
recently been the subject of many experimental, theoretical and
simulational studies
\cite{Fenistein03,Fenistein04,Luding04,Unger04a,Fenistein06,Cheng06,Depken06,Depken06b,Unger07,Torok07}.
In experiments the cell has cylindrical form called the modified Couette
cell \cite{Fenistein04}. It is a container whose bottom is divided into a
central disk and an outer ring. The disk rotates slowly with respect to the
rest of the container. When sand is filled in, it is dragged along by
the rotating central bottom disk so that a shear zone emerges 
in the material, where
the shear deformation is localized. It starts at the perimeter of the
bottom disk, spreads into the bulk and reaches the top surface, if the
filling height is not too large. The shear zone can be characterized by
its central sheet (the sheet of the maximum shear rates)
and the width of the zone around the central sheet.

Depending on the experimental conditions the behavior of the central sheet
can be quite complicated. Due to the cylindrical shearing it gets a
nontrivial curved shape with decreasing radius towards the top of the
system. The shape depends strongly on the filling height
\cite{Fenistein04}. For large $\Hfill$ the central sheet even detaches
from the top and dives entirely into the bulk
\cite{Unger04a,Cheng06,Fenistein06} forming a cupola-like shape. If two
materials are used, the central sheet can be refracted, when the
shear zone leaves one material and enters into the other \cite{Unger07}.
All these effects will be avoided hereafter in the paper. We deal
with the linear version of the split-bottom cell (see later)
\cite{Depken06,Depken06b}, where the central sheet 
remains a vertical plane. Therefore the flow becomes simpler and widening
of the shear zone can be analyzed more easily.

The width of the shear zone $W$
has been found to be an increasing function of the bulk height $z$ and also
an increasing function of the filling height $\Hfill$
\cite{Fenistein04,Cheng06,Torok07}. The width at the top
($W_\text{top}$) grows more slowly than the filling height but faster than
the square root of ${\Hfill}$. The experimental data suggest that
$W_\text{top}$ is approximately a power law with exponent $2/3$
\cite{Fenistein04,Fenistein06}.

Although it is a very basic question, where this type of rheology comes from,
no satisfying description has been found so far. There are proposals
what the governing mechanism of shear zones could be. Some of these
theoretical approaches \cite{Depken06,Depken06b,Unger04a,Torok07} are
reassuring in the sense that they lead to wide shear zones and, at the same
time, satisfy the condition of rate independence. One approach
\cite{Unger04a,Torok07,Unger07} is based on the weakest sliding surface which
fluctuates during the flow, another one \cite{Depken06,Depken06b} is
based on the variation of the effective friction coefficient depending on
the orientation of the local shear plane. At the current stage these models
are not very well established and concerning the details they leave many
questions open. 

It is hard to refine existing models or propose new
candidates because not enough details are known about the flow. Especially,
precise data are needed that are measured in the bulk, regarding
e.g. velocity and stress fields.

With our present study we provide some new details about the flow in the
split-bottom shear cell. We perform DEM simulations where the velocities
and stresses are easily accessible in the bulk. Previous experiments and
simulations were in gravity which leads to an inhomogeneous pressure
distribution. We analyze shear zone formation also in a 
zero gravity environment in order to clarify the role of
gravity.

\section{Description of the numerical experiment}

\begin{figure}
\includegraphics[scale=.5]{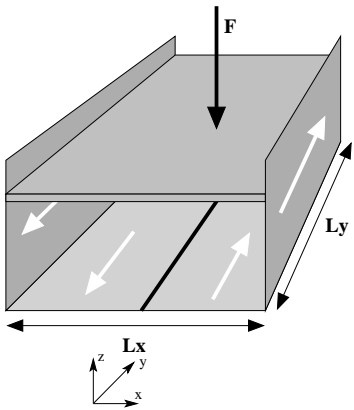}
\caption{\label{fig-shearcell} The linear split-bottom cell.}
\end{figure}

In our simulations we examine a linear version of the split-bottom shear
cell \cite{Depken06,Depken06b} shown in Fig.~\ref{fig-shearcell}. Here the
bottom is cut along a straight line. The left and the right sides of the
boundary move along the $y$ axis in opposite directions both with velocity $v_\text{shear}$. 
In $y$-direction periodic boundary conditions are applied. Small grains are
glued to the side walls and to the bottom in order
to make their surface rough.

We have a frictionless piston on the top of the system. Its position in
$z$-direction gives the filling height of the system. The piston has a
compressing force $F_\text{pist}$ on it acting in negative
$z$-direction. We use
two ways to put the system under pressure. Either we apply a large force
$F_\text{pist}$ and set gravity to zero or we use gravity instead and
put only a
weak force on the piston. The role of the piston in the former case is to
provide the confining pressure on the system. In the latter case it keeps
only the top surface flat and ensures a constant filling height for the
whole system. Then the piston has negligible effect on the pressure
distribution in the bulk which is generated essentially by gravity.

Our simulations are discrete element simulations based on the method of
contact dynamics \cite{Jean99,Brendel04}. The grains are noncohesive, rigid
and spherical interacting via frictional contact forces. The value of the
friction coefficient is set to $0.2$.
Throughout this paper every length is measured in units of the
maximum grain radius. Radii are uniformly distributed between $0.8$ and
$1.0$.

We tested various system sizes.
The number of the grains $N$ contained by the shear cell varies between
$1\,000$ and $100\,000$. The width $L_x$, the length $L_y$ and the filling
height $\Hfill$ of the systems range from $20$ to $240$, from $12.5$ to
$75$, and from $8$ to $70$, respectively.

Our simulation corresponds to an experimental situation where the grains
have density $2400$ $kg/m^3$ and maximum radius $1$ $mm$. The value of
$v_\text{shear}$ is set to $0.7$ $cm/s$ (unless stated otherwise). The force
$F_\text{pist}$ is chosen proportional to the surface of the piston in
order to maintain the same pressure. This pressure is $500$ $N/m^2$ when
gravity is switched off. Together with gravity the pressure on the piston
is set to $25$ $N/m^2$.

The preparation of the system starts from a gas state where grains have
random positions. First we compactify
the material with the piston then gravity is switched on if needed and the
shearing starts. Before measuring velocities and stresses we let the
system relax in order to reach stationary flow.

\section{Results}

\subsection{Orthogonal velocities}

First we examine whether the shear cell generates any convection orthogonal
to the shear direction $y$. The components of the coarse-grained velocities
$v_x$, $v_y$ and $v_z$ are functions of the coordinates $x$ and $z$ (the
coordinate $y$ is averaged out). In the present shear cell $v_x$ and $v_z$
would vanish for a laminar flow of a Newtonian fluid, however, this does
not hold a priori for quasi-static flow of granular media. One could imagine
various kinds of stationary flows with non-vanishing convection in
$x$-$z$-plane, e.g., where grains,  besides moving in $\pm
y$-direction, slowly rise near to the symmetry plane and descend far away
from it. 

\begin{figure}
\includegraphics[scale=.6]{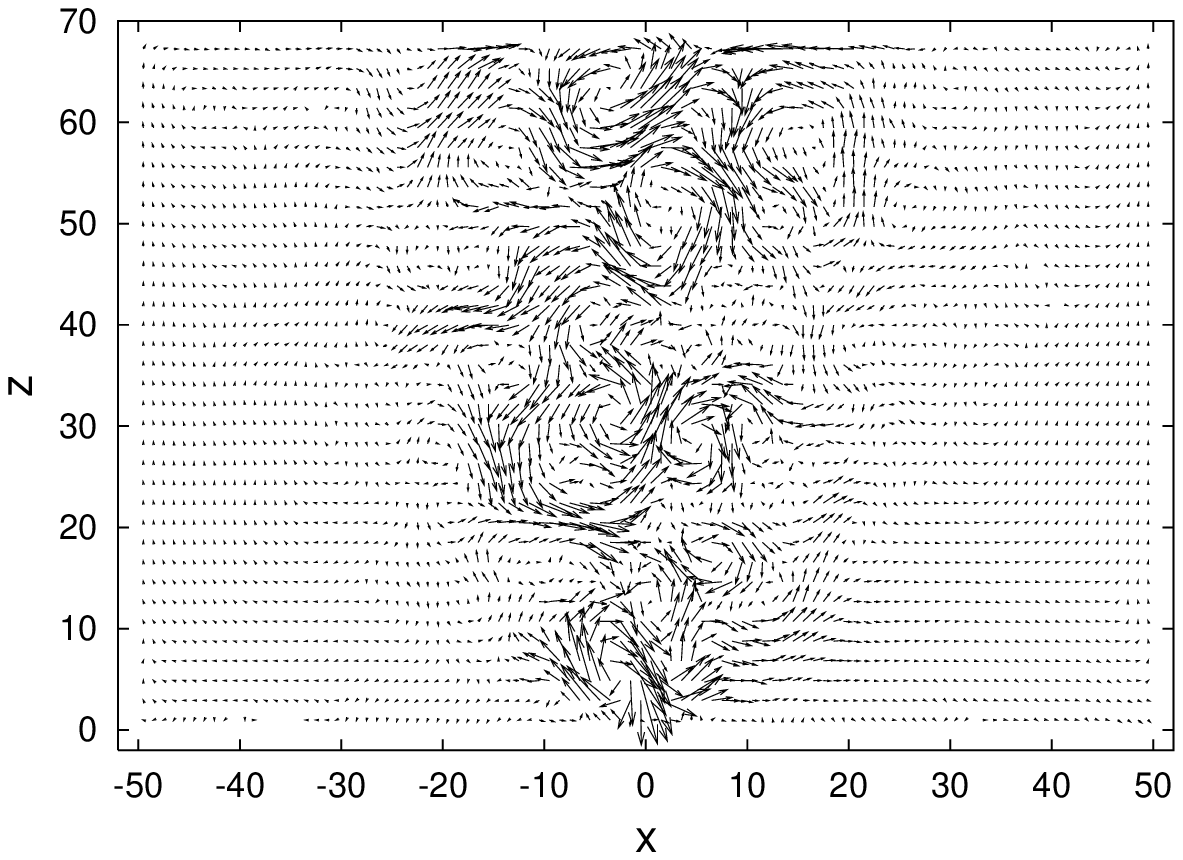}
\includegraphics[scale=.6]{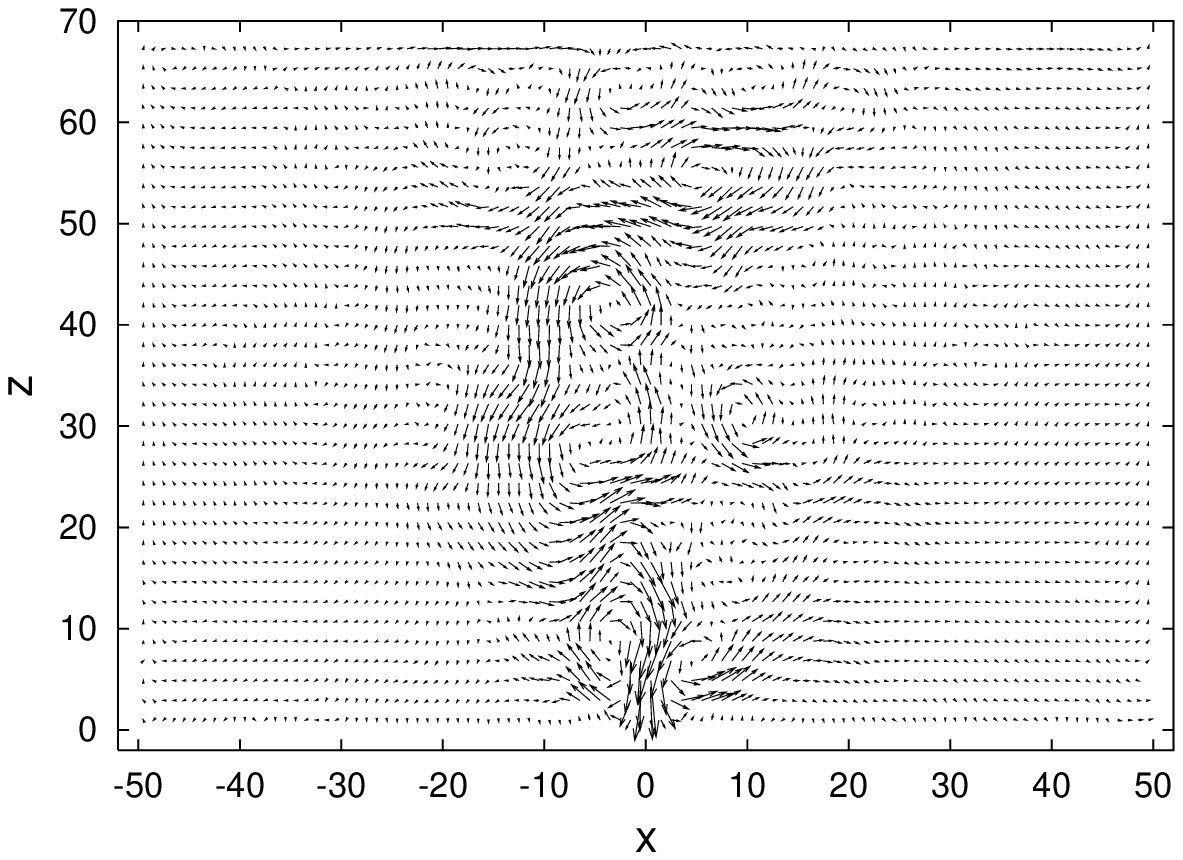}
\caption{The average velocity field in the cross section of the cell
  orthogonal to the shear direction. The time average is taken over a shear
  displacement $60$ for (a) and $300$ for (b).}
\label{orthovelo}
\end{figure}

Fig.~\ref{orthovelo} shows a typical orthogonal velocity field.
This simulation contains $100000$ grains and is
performed without gravity. Before recording velocity data we sheared the
system for a long time in order to achieve a steady state. During this
pre\-shearing the system had total shear displacement $\lambda=500$ (this is the
displacement of the two sides of the shear cell with respect to each
other). The velocities shown in Fig.~\ref{orthovelo}.a are obtained by an
average over a further shear displacement $60$. This velocity field seems
to be merely random fluctuation where vortices arise due to mass
conservation. The largest velocities are located near the symmetry plane
of the cell and their magnitude is about $200$ times smaller than the shear
velocity of the cell. These velocities decrease further if the average is
taken over 
larger shear displacements. This is shown in Fig.~\ref{orthovelo}.b where
the shear displacement is $5$ times larger compared to
Fig.~\ref{orthovelo}.a.  Thus we find no convection orthogonal to the shear
direction. If such convection is induced by the shearing then it must be at
least $4$ orders of magnitude smaller than the shear velocity itself.

In the followings we will focus on the motion along the shear direction $y$.

\subsection{Velocity profiles}

The shear flows found in our simulations are in agreement with previous
experimental and numerical measurements in the modified Couette cell
\cite{Fenistein03,Fenistein04,Fenistein06,Cheng06,Depken06b}. Our new
results concern the slow 
evolution in the outskirts of the shear zone, its widening with
increasing distance from the bottom slit (Sec. C), and the influence
of gravity (Sec. D). In particular it will be shown 
that gravity has surprisingly little effect on the properties of the
shear zone.

\begin{figure}
\includegraphics[scale=.6]{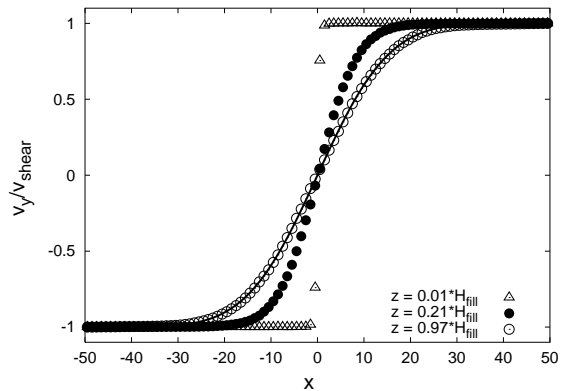}
\caption{\label{3profiles} The velocity profiles are taken from the same
  system at three different heights $z$. The line shows an error
  function fit.}
\end{figure}

In order to examine the velocities $v_y$ we divide our system into
different slices at constant heights $z$. $z$ ranges from zero to the
filling height $\Hfill$. At each height $z$ the velocity $v_y$ goes from
$-v_\text{shear}$ to $v_\text{shear}$ as $x$ is increased. This transition
is very sharp at the bottom, where the boundary condition prescribes a step
function, and broader towards the top of the system. Fig.~\ref{3profiles}
shows the profiles for a system without gravity in several heights.

The velocity profiles can be well fitted
with error functions \cite{Fenistein04}. Consequently, the shear rate
$\dot\gamma_{xy}$ as a function of $x$ is a Gaussian curve. We define the
\emph{width of the shear zone $W(z)$} as the square root of the second
moment of the (normalized) shear rate at height $z$ thus the width of the
zone equals to the width of the corresponding Gaussian curve.

\begin{figure}%
\includegraphics[scale=.6]{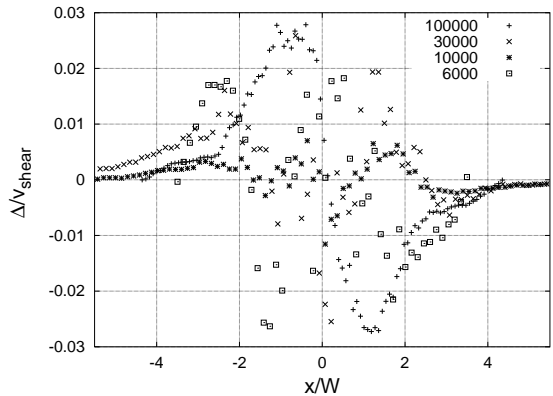}
\caption{\label{fitcomp1} The difference $\Delta$ between the velocity data
  and the fit by error function is shown. The data are recorded in nine
  different systems. Total number of grains used in the
  simulations are between $6\,000$ and $100\,000$. Minimum and maximum
  heights are $25.6$ and $69$.
  In all cases shown here gravity is set to zero. Only the fits of the
  velocity profiles at $z = \Hfill$ are evaluated.
  For each system the coordinate $x$ is normalized by the width of the shear
  zone measured in the top layer.}
\end{figure}

\begin{figure}%
\includegraphics[scale=.6]{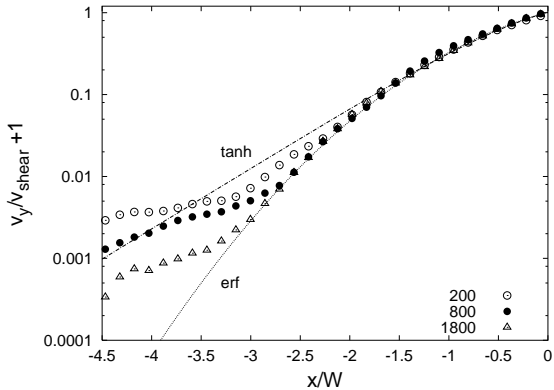}  
\caption{\label{fitcomp3}%
Slow time evolution of the tail of the velocity profile
is shown for a system of $10\,000$ grains at the filling height
$\Hfill=25.6$. For
the three different profiles the measurements are started at shear
displacement $200$, $800$ and $1800$. Each profile represents an average
over an additional shear displacement $200$. The two lines are fit curves
of the velocity data: the upper one is a hyperbolic tangent, the lower one is
the error function. It can be seen that the velocities follow the
Gaussian tail of the error function the better the larger the shear
deformation of the sample is.}
\end{figure}

The accuracy of the fit by an error function is assessed in
Figures~\ref{fitcomp1} and \ref{fitcomp3}. In Fig.~\ref{fitcomp1} the
deviations between data and fit are plotted for several
systems. Deviations are random and approximately $2\%$ of $v_\text{shear}$
near the center. Further away from the center the errors become
smaller, however, systematic deviations can be seen: They are positive on
the left and negative on the right hand side. From Fig.~\ref{fitcomp3} one
can conclude that these systematic deviations are going to vanish, if
one lets the simulations run longer. The velocity profile approaches
the error function shape first in the center, but much more slowly in
the outskirts. If we let the simulations go on, the tail of the
velocity profile keeps evolving and is getting
closer and closer to the Gaussian tail of the error function. 
We will come back to the relaxation process in section \ref{criticalrelax}.

The above properties of the velocity distribution are in agreement with
experimental data \cite{Fenistein04} which have been achieved in gravity in
a modified Couette cell, whereas the simulation data presented in
Fig.s~\ref{3profiles}, \ref{fitcomp1} and \ref{fitcomp3} were obtained
for zero gravity. As we will discuss below, gravity has indeed only very
little influence on the velocity profile.

\subsection{Widening of the shear zone in the bulk}

Previous experimental and numerical studies revealed that the shear zone
becomes wider as it goes from the bottom $z=0$ towards the top
$z=\Hfill$. There are some experimental and theoretical indications
reported in \cite{Depken06} that
$W(z)$ may be a power law with exponent between $0.2$ and $0.5$.
However, we are not aware of any conclusive experimental data concerning
the exact shape of the function $W(z)$. It is a crucial question what the
functional form is because it provides a very strong test for
theories. Such tests are 
clearly needed as the problem of quasi-static flow and shear zone formation
is far from understood. This field is in the stage of searching for
candidates of models in order to gain a better understanding of basic
phenomena. 

\begin{figure}%
\includegraphics[scale=.46]{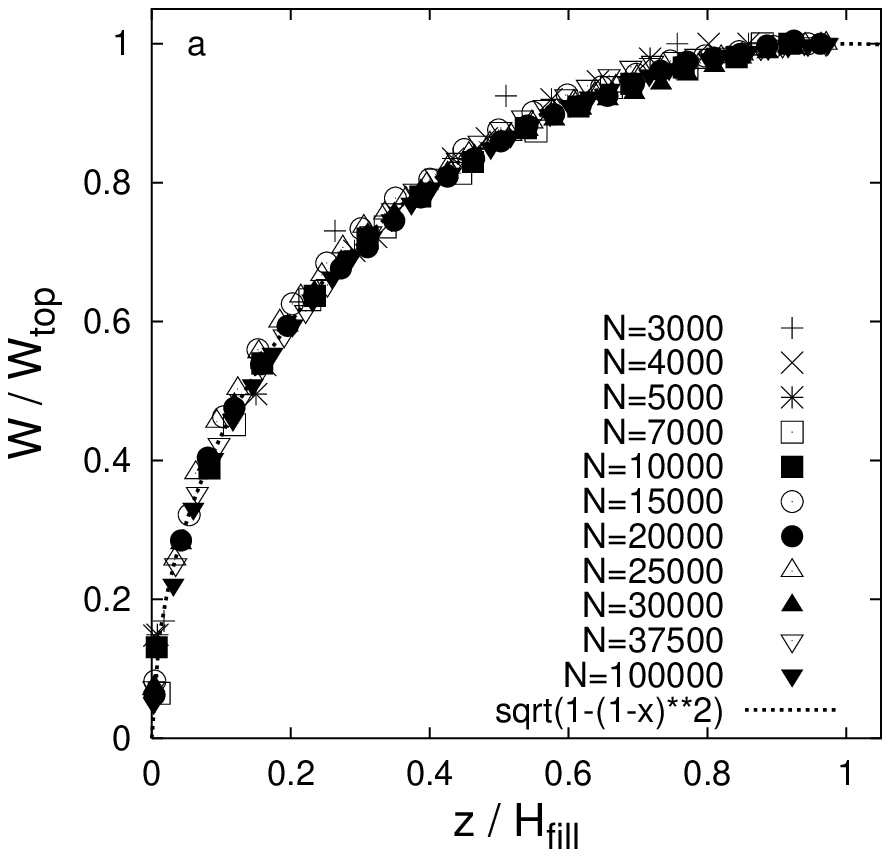}
\includegraphics[scale=.46]{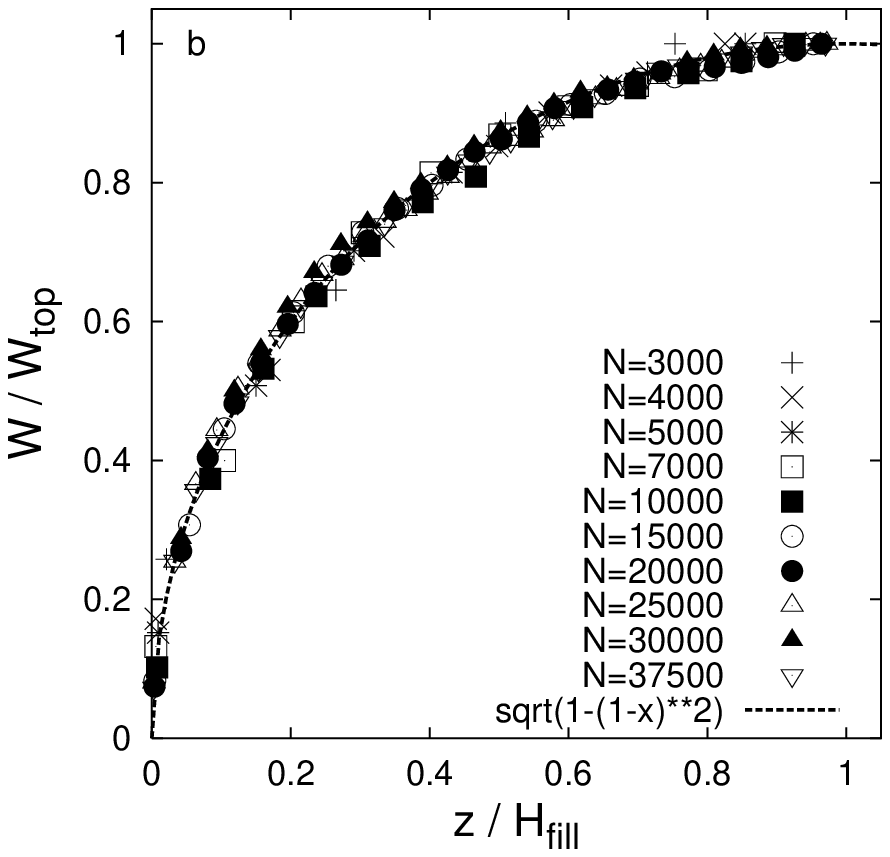}
\caption{\label{bulkwidth1} Data collapse of width - bulk height data is
  achieved when rescaled with the maximum width and the filling height,
  respectively. Systems containing different number of grains $N$ are
  plotted here. Filling heights ranges from $9$ to $69$. a) without
  gravity, b) with gravity.}
\end{figure}

It is a nice feature of computer simulations that one can easily access the
velocity data also inside the bulk. Based on these data we are able to
deduce the functional form of $W(z)$. We tested many systems with
different filling heights. It turns out that all the width data
collapse if plotted in the frame $(W(z)/W_\text{top}\, , \, z/\Hfill
)$. The collapse of the data can be seen in Fig.~\ref{bulkwidth1}. The
master curve that the data follow is a quarter of a circle:
\begin{equation}
  W(z) = W_\text{top} \sqrt{1-\left(1-\frac{z}{\Hfill}\right)^2} \, .
  \label{circle}
\end{equation}
Thus we find that the widening of the shear zone starts with an exponent
$1/2$ for small values of $z$ but soon departs from the power law. $W(z)$
hits the top of the system at a right angle. 

\begin{figure}%
\includegraphics[scale=.47]{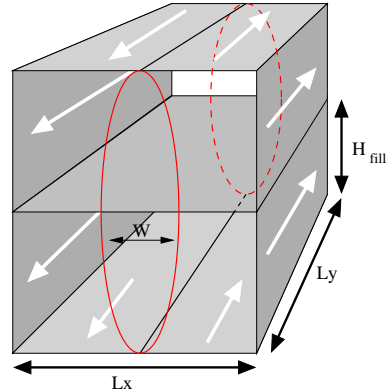}
\caption{\label{shearmirror} The system and its mirrored counterpart
  above.}     
\end{figure}

This latter condition of the right angle seems to be quite reasonable at least
for the
case when gravity is switched off. The frictionless piston we apply
at the top exerts no drag force on the material, but only applies normal
pressure on the system. An equivalent situation can be achieved
if we take the original system together with its mirror image (see Fig.~\ref{shearmirror})
and at the same time we leave the piston away.
Then we have a split boundary both at the bottom and at the top. The total
height of the system is then two times the original filling height. For
symmetry reasons there is no drag force between the upper and lower parts
of the system which explains the equivalence. And again for symmetry
reasons the curve $W(z)$ must be perpendicular to the plane of the removed
piston.

Interestingly the presence or absence of gravity has no influence on the
data collapse: the master curve given by Eq.~\ref{circle} is valid for both
cases (Fig.~\ref{bulkwidth1}). 

\subsection{Role of gravity}

Significant efforts have been made recently to understand the behavior of
wide shear zones. However, all experimental, theoretical and numerical
studies subjected to split bottom shear cells (either linear or cylindrical
cells)
\cite{Fenistein03,Fenistein04,Luding04,Unger04a,Fenistein06,Cheng06,Depken06,Depken06b,Torok07}
investigated shear zones under gravity. 

Gravity leads to an inhomogeneous stress distribution in the system. Stresses
even go to zero as the free surface of the sample is approached. It is
not unplausible to imagine that gravity might be responsible for
certain features of the shear zones (e.g. their widening towards the
free surface).

\begin{figure}%
\includegraphics[scale=.6]{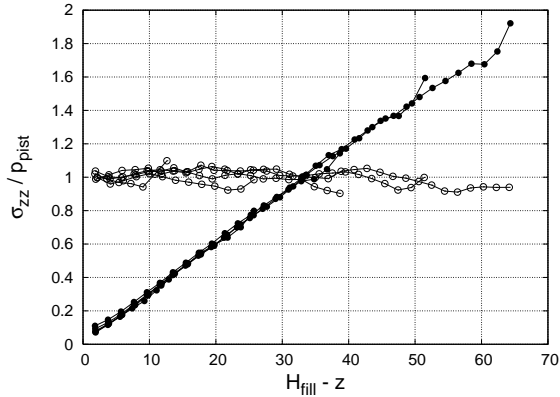}
\caption{\label{sig_zz} The stress $\sigma_{zz}$ as the function of
 depth. Full circles are recorded in gravity while open circles correspond
 to zero gravity where the entire pressure is provided by a
 piston. For each case we plot $4$ different curves which represent
 different number of grains filled  
 in the shear cell ($N$: $5000$, $10000$, $15000$, $20000$ and
 $25000$). The filling height $\Hfill$ changes proportionally to
 $N$. In gravity $\sigma_{zz}$ corresponds to hydrostatic
 pressure. $p_\text{pist}$ is the pressure on the piston in zero gravity.}
\end{figure}

In Fig.~\ref{sig_zz} we show the stress component $\sigma_{zz}$ at the
symmetry plane of the shear cell. It can be seen that $\sigma_{zz}$ 
is proportional to the depth in case of gravity and approximately constant
without gravity.

Recently, Depken et al. \cite{Depken06,Depken06b} argued that in the
quasi-static regime one cannot achieve wide shear zones, if the
effective friction coefficient $\mu_\text{eff}$ is assumed to be
constant. For constant $\mu_\text{eff}$ the shear zone should localize
to a thin layer. In their model the widening of the shear zone is
attributed to the dependence of $\mu_\text{eff}$ on the angle
$\Theta$ between the direction of gravity and
the local tangent plane of the constant velocity surfaces. (In other
words they assumed that the frictional properties of the material
depend on the orientation of gravity with respect to the local sliding
plane.)

Our simulation data do not support the above picture. The effective
friction $\mu_\text{eff}$ might vary throughout the shear zone,
however, the direction of gravity does not seem to play any important
role here. As we discussed in previous sections the velocity profiles
are qualitatively the same no matter whether gravity is present or
not. Surprisingly, body forces or decreasing pressure are not needed
for wide shear zones.

\begin{figure}%
\includegraphics[scale=.6]{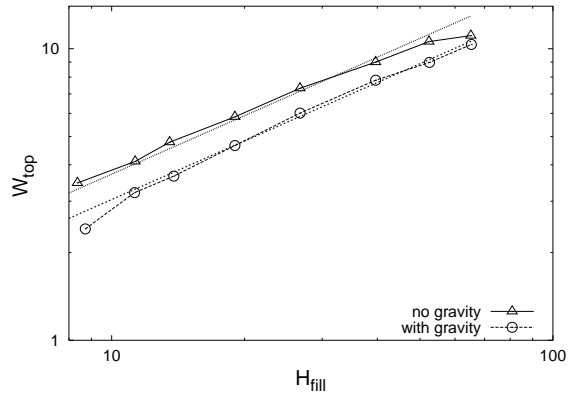}
\caption{\label{WH} Width of the shear zone at the top of the systems
  plotted as function of the filling height. Gravity reduces the width of
  the shear zones. Lines are not fits, they only show the slope of
  exponent $2/3$.} 
\end{figure}

In fact, shear zones exhibit even larger width when gravity is
switched off. The width gets larger by a factor $1.2 \pm 0.1$. This
can bee seen in Fig.~\ref{WH} where the top width of the shear zone
for various filling heights is plotted.

Why gravity contracts the shear zone can be explained by the
pressure distribution. At the bottom the pressure, and thus also the
frictional forces, are much stronger than at the top of the
system. Therefore the rheology is dominated by the lower part of the
system. Compared to the case of homogeneous pressure this acts as if the
system had effectively smaller filling height which leads to a smaller
width of the shear zone. The contracting effect of gravity is also
discussed in \cite{Ronaszegi07} based on the principle of minimum
energy dissipation.

The experimental data of Fenistein et al.
\cite{Fenistein04,Fenistein06} showed that $W_\text{top} (\Hfill)$ is
approximately a power law with exponent $2/3$. $W_\text{top} (\Hfill)$
found in our simulations is shown in a log-log plot in Fig.~\ref{WH},
where the exponent $2/3$ is also indicated for comparison. The data
follow approximately the experimental behavior. For a precision value
of the exponent or a discussion of deviations from a power law better
statistics is needed, however.

\subsection{Influence of additional parameters}

It was assumed so far that the flow is quasi-static, i.e. the
shear velocity $v_\text{shear}$ is small enough that no rate-dependence is
observed in the behavior of the shear zone. We also intended to
choose the width $L_x$ and length $L_y$ of the systems large enough in
order to exclude their influence on the flow. Furthermore, it was assumed
that the analysis of the shear zone was taken after initial transients in
the stationary flow regime, i.e. the total shear displacement $\lambda$ was
large enough to ensure complete relaxation.

In this section we check the influence of the parameters $L_x$, $L_y$,
$\lambda$ and $v_\text{shear}$ to show that they are chosen properly
and do not alter the properties of the shear zone. For this purpose we
take one of the previous samples as reference system and demonstrate
the role of the four parameters there. The reference system contains
$N=10\, 000$ grains, has the size $L_x=80$, $Ly=25$, $\Hfill=25.6$
(this system appeared already in Figures \ref{fitcomp1},
\ref{fitcomp3}, \ref{bulkwidth1} and \ref{WH}).

\begin{figure}%
\includegraphics[scale=.45]{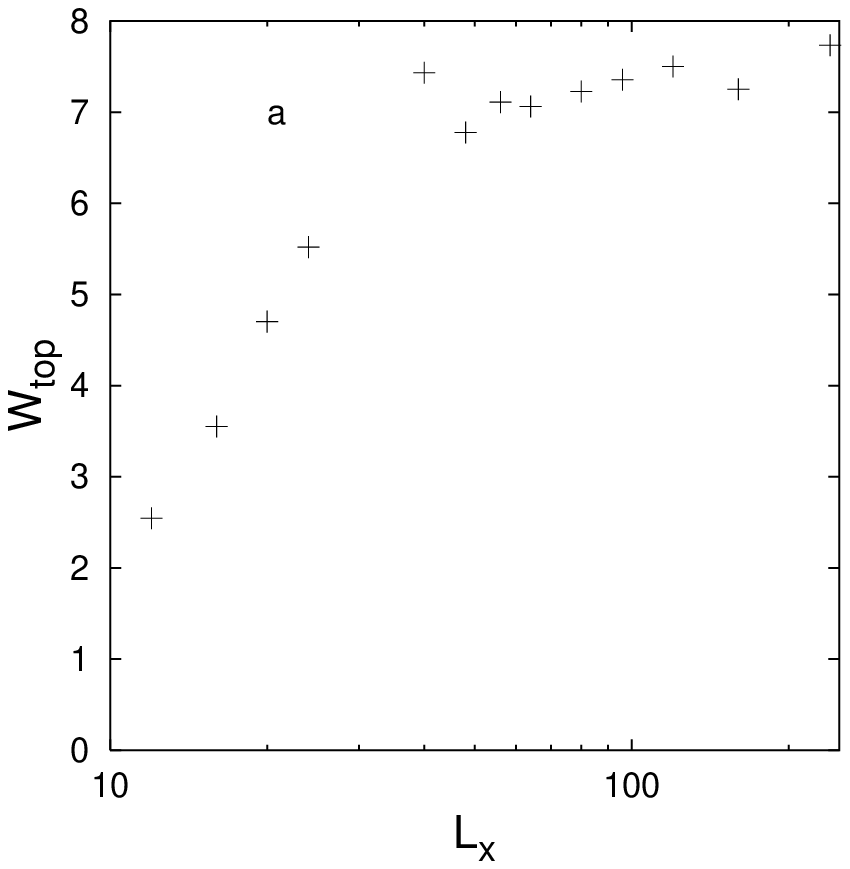}
\includegraphics[scale=.45]{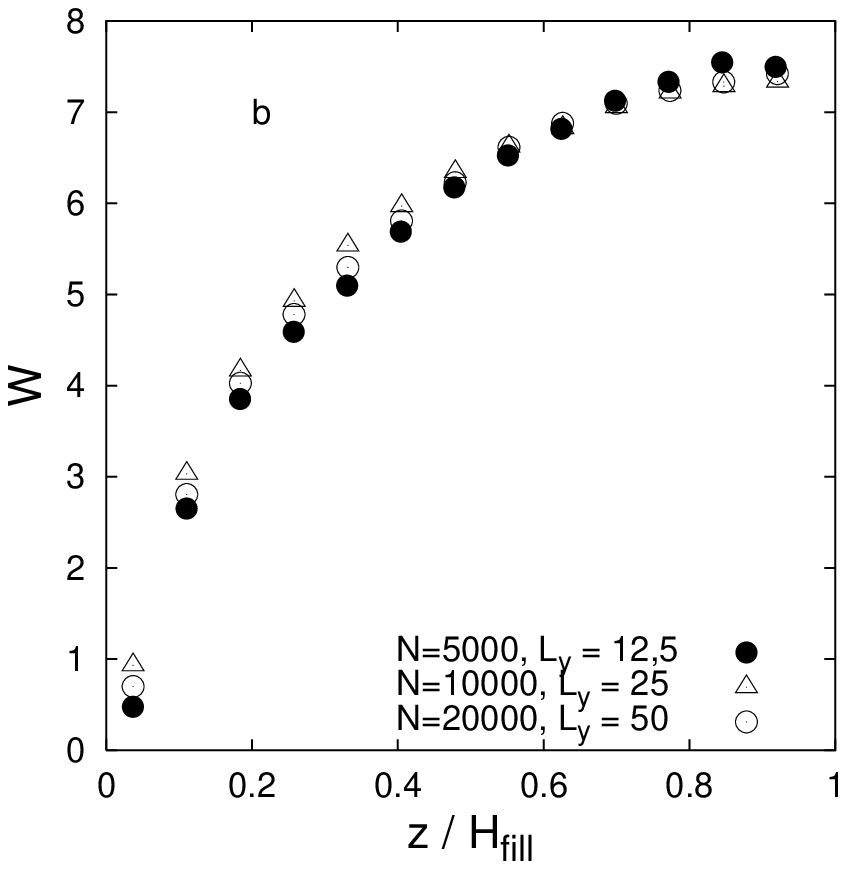}\\
\includegraphics[scale=.45]{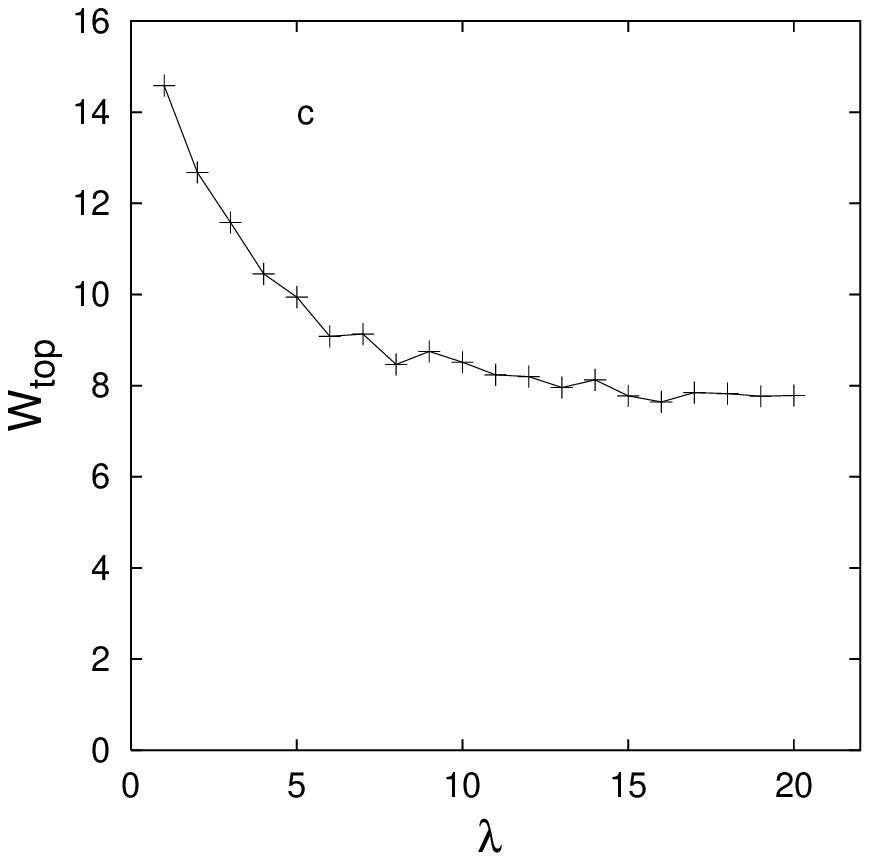}
\includegraphics[scale=.45]{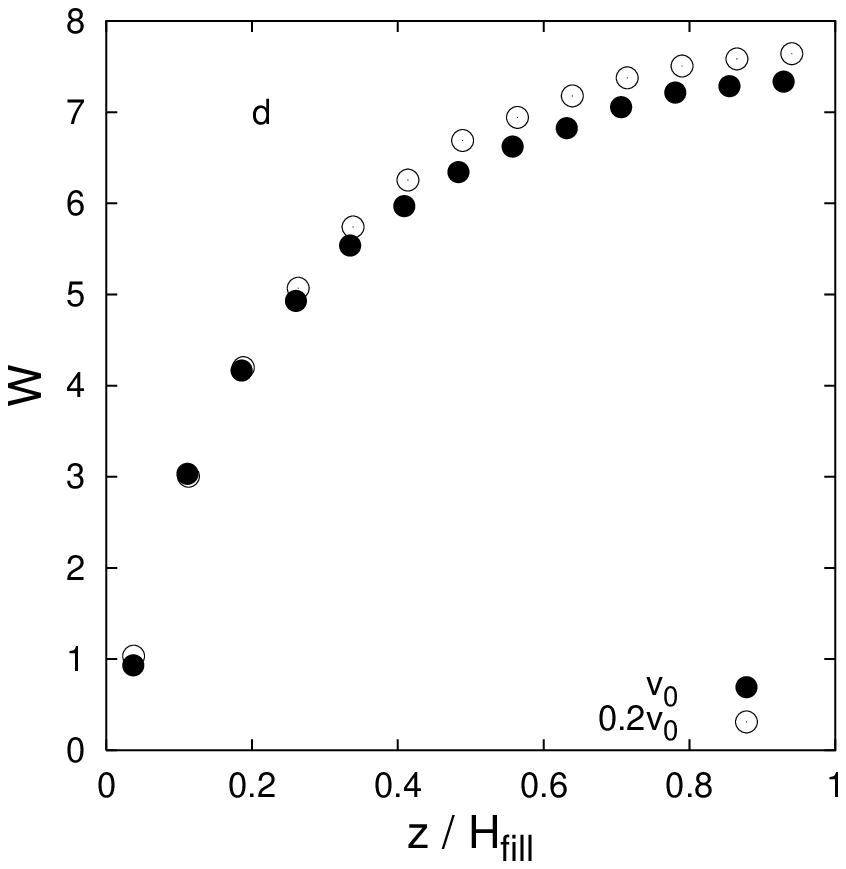}
\caption{\label{check}
Robustness of the width of the shear zone. a) The effect of the system
width $L_x$ on the top width of the shear zone. b) The width of the shear
zone is plotted as function of the bulk height $z$ for three different
lengths $L_y$ of the shear cell. c) Relaxation of the top width as function
of the shear displacement $\lambda$. d) The influence of
the shear velocity $v_\text{shear}$ on the curve $W(z)$. Full circles
denote five times larger shear velocity than the open circles.
}
\end{figure}

First we vary the width of the system $L_x$ and the number of grains $N$
proportionally to $L_x$ in order to maintain the same filling height. Other
parameters are constant. The effect on the width of the shear zone can bee
seen in Fig.~\ref{check}.a. Around the reference point $L_x=80$ the width
remains constant. The shear zone ``feels'' the effect of the side walls
only if $L_x$ drops below $40$ where the shear zone can be strongly
contracted by lowering $L_x$.

For the parameter $L_y$ we test values $12.5$, $25$ and $50$ (again $N$ is
changed proportionally) while other parameters are
constant. Fig.~\ref{check}.b shows that the width of the shear zone remains
essentially the same (changes are small and without systematics).

In order to avoid initial transients all the systems presented in this
paper undergo a preshearing over a total displacement of more than
$\lambda > 60$, and only after that we start collecting data.
Fig.~\ref{check}.c shows that the transient period is indeed completed
during the preshearing: The width of the shear zone becomes
independent of the displacement $\lambda$. The origin of the initial
transient is discussed in Sec.~\ref{criticalrelax}.

Due to the larger time resolution applied here velocity fluctuations become
also larger. In order to improve the statistics $20$
parallel simulations were taken. In each case the system had the same
macroscopic parameters as our reference system. In Fig.~\ref{check}.c the
time evolution of 
$W_\text{top}$ was achieved by ensemble average over the $20$ simulations.

Next we turn to $v_\text{shear}$.
If pressure conditions are the same it depends on the speed of driving
whether the flow is quasi-static. For relative large driving speeds, but
still in the dense flow regime, inertia effects come into play which
weaken shear localization and results in wider shear zones. Then the width
of the shear zone can be made smaller by lowering the driving speed. But if
the driving gets slow enough the rheology becomes independent of the
driving rate and the shear zone reaches its minimum width.
We test the effect of the driving rate in our reference system by
taking five times smaller shear velocity. The new bulk-width of the shear
zone is compared with the original one in Fig.~\ref{check}.d. The data show
no further decrease of the width. The width for the reduced
driving is even slightly larger due to random fluctuations. Within the
accuracy of our numerical measurement the two curves can be regarded
as equal.

\subsection{Shear zone versus critical zone}
\label{criticalrelax}

The shear zone is the region where the major part of the shear deformation
takes place. It is described by the function $W(z)$. The material, however,
is not solid outside the shear zone either. Split-bottom
cells fluidize the material everywhere, however the shear rate becomes many
orders of magnitude smaller far from the shear zone.

In this section we would like to discuss the concept of the critical
state. It has not got any attention so far in the context of the
split-bottom shear cells, although, it leads to the emergence of a relevant
and new type of zone.

It is known that the mechanical properties of granular media are influenced 
strongly by the preparation. If one starts shearing a packing the behavior
can be different depending on the initial state (density, structure of the
contact network, etc): It can lead to different stress responses,
effective frictions, dilation or contraction, etc. However, if the material
experiences large enough local strain it reaches a unique state regardless
of the preparation history. This is the critical state
\cite{Wood90,Radjai04,Craig04,Cruz05,Kadau05} where the material organizes
and maintains its 
microscopic inner structure on shearing. After the critical state has been
reached unlimited shear deformation can occur without changes of stresses
or density. The characteristic deformation scale needed to erase the memory
of the material and reach the critical state is typically around $\gamma =
0.2$, where $\gamma$ is the cumulative shear strain.

Before the shearing starts in the split-bottom cell the fabric of the
material reflects the direction of the initial compression or
gravity. With the shear deformation this structure is destroyed and new
contacts are created against the direction of the shear. This gives rise to
strain hardening: the resistance of the material against shear is
increased. This does not happen simultaneously all over the sample. When
regions in the middle of the shear cell are already in the critical state,
regions far away can still be frozen in the initial configuration.

At the beginning, the zone of the critical state starts growing from the
split line at the bottom. It reaches quickly the top of the system and also
spreads sideways. As the shear rate is very small far from the symmetry
plane, the growth of the critical zone becomes extremely slow here. 
In that sense a steady state can not be reached in the whole system. One
expects the flow to become stationary only inside the critical zone.

The growth of the critical zone comes into sight in
Fig.~\ref{fitcomp3}. The velocity of a given point in the system reaches
its final value only after the region becomes critical. The position where
the velocity data depart from the stationary curve mark the border of the
critical zone. The different velocity profiles recorded in different stages
of the simulation show how the width of the critical zone is increasing with
time. It can be attributed to the strain hardening why the shear rate is
reduced at a given position after it becomes critical.

The reduction of the shear rate due to strain hardening also explains the
relaxation of the shear zone that is shown in Fig.~\ref{check}.c. In the
early stage of the simulation where the critical zone is smaller than the
shear zone the shear rate is slightly enhanced outside the critical
region. This makes the shear zone wider a little bit. After the whole shear
zone becomes critical this additional widening effect ceases, and
$W_\text{top}$ is reduced to its final value.

According to this interpretation the initial transient of the shear zone
is due the time evolution of the critical region. One can estimate
the shear displacement $\lambda$ that corresponds to the transient
period by matching the size of the stationary shear zone and the growing
critical zone. Using the stationary velocity profile and the cutoff shear
strain $\gamma = 0.2$ one gets $\lambda=7.7$  for the transient. This is
in excellent agreement with the relaxation observed in Fig.~\ref{check}.c.

The growth of the critical zone and the transient of the shear zone can be
reproduced within the framework of a simple lattice model
\cite{Ronaszegi07} developed for quasi-static shear flows.

The presence of the critical zone can be observed also in the stress
field. The next section is subjected to this question.

\subsection{Stresses}

As the average local velocities have only a $y$-component ant the
system is translational invariant in $y$-direction, the local strain
rate in the $(x,y,z)$-frame has the form
\begin{equation}
\frac{1}{2}\left(\begin{array}{ccc}
0               & \frac{\partial v_y}{\partial x} &       0\\ 
\frac{\partial v_y}{\partial x}&0&\frac{\partial v_y}{\partial z}\\        
0               & \frac{\partial v_y}{\partial z} &       0              
\end{array}\right)\ .
\end{equation}
By means of a rotation around the $y$-axis a new local frame $(u,y,v)$
can be chosen in which the strain rate is
\begin{equation}
\frac{1}{2}\left(\begin{array}{ccc}
0               & \frac{\partial v_y}{\partial u} &       0\\ 
\frac{\partial v_y}{\partial u}&0& 0\\        
0               & 0  &       0              
\end{array}\right)\ .
\end{equation}

If initial conditions are forgotten in the critical state (also called
steady state flow), stress and fabric tensors are expected to have the
same principal axes as the strain rate tensor. Then the stress tensor
must have the form ~\cite{Depken06}  
\begin{equation}
\begin{pmatrix}
\sigma_{uu}&\sigma_{uy}&\sigma_{uv}\\
\sigma_{yu}&\sigma_{yy}&\sigma_{yv}\\
\sigma_{vu}&\sigma_{vy}&\sigma_{vv}
\end{pmatrix}
=
\begin{pmatrix}
P&\tau&0\\
\tau&P&0\\
0&0&P'
\label{eq-stress}
\end{pmatrix}\ .
\end{equation}

Depken et al.~\cite{Depken06b} tested stresses in the linear
split bottom cell by soft particle molecular dynamics simulations. They
found the expected behavior in the middle part of the cell where stress 
and strain tensor were co-linear and stresses took the form (\ref{eq-stress}).

\begin{figure}%
\includegraphics[scale=.7]{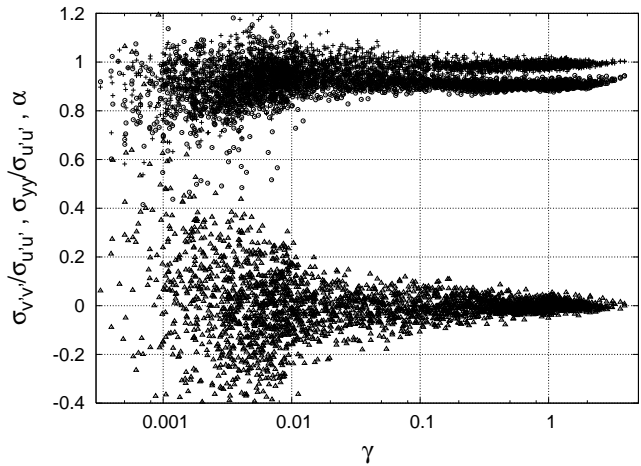}
\caption{\label{stress-ratios}
The effect of the local shear strain $\gamma$. Crosses and circles show
ratios of normal stress components $\sigma_{yy}/\sigma_{u'u'}$ and
$\sigma_{v'v'}/\sigma_{u'u'}$, respectively. The parameter $\alpha$ (triangles)
indicate the angle between the local shear stress and shear strain.}
\end{figure}

\begin{figure}%
\includegraphics[scale=.7]{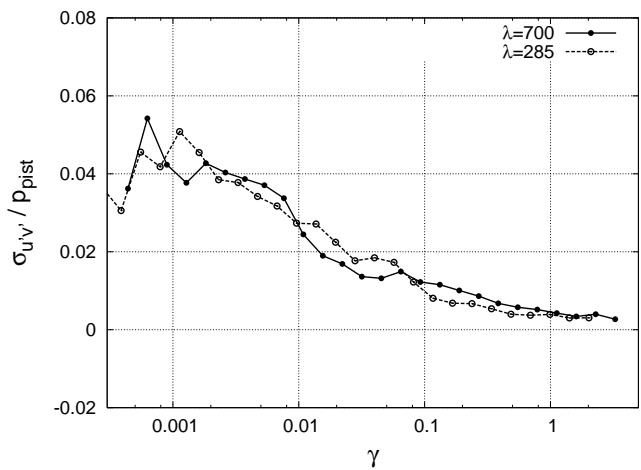}
\caption{\label{remainingShearstress}
The remaining shear stress $\sigma_{u'v'}$ vanishes for large shear
deformations. The open circles and dots correspond to measurements started
at $\lambda=285$ and $700$, respectively. In both cases the measurement
lasted over a period $\lambda=120$.}
\end{figure}

However, if the material still remembers its initial structure the
alignment of stress and strain is not necessarily valid.
Therefore we determine a second
local frame $(u',y,v')$ from the condition 
$\sigma_{yv'}=\sigma_{v'y}=0$ and evaluate the angle $\alpha$ between
the two directions $(u,v)$ and $(u',v')$ in the $(x,z)$-plane. 
The lower part of Fig.~\ref{stress-ratios} shows indeed that the angle
$\alpha$ approaches 0 for large shear deformations $\gamma$, while the 
principal directions of stress and strain rate tensor differ during
the transient. Correspondingly, $\sigma_{u'v'}$ approaches 0 for large
shear deformations (see Fig.~\ref{remainingShearstress}), as predicted
by Eq.\ref{eq-stress}.

Our numerical test was based on another method than the one used by Depken et
al.~\cite{Depken06b}, the contact dynamics
algorithm \cite{Jean99,Brendel04}, and we used slightly different
conditions (zero gravity, piston). Nonetheless, we found the same
behavior for regions where the material experienced large shear
deformation. Here stress and strain tensors align
and the stress corresponds to the reduced form in Eq.~\ref{eq-stress}.

Stress data recorded in a system of $100\,000$ grains with total shear
displacement $820$ are presented here as the function of
the cumulative local shear strain $\gamma$. Fig.~\ref{stress-ratios}
shows stress 
ratios $\sigma_{yy}/\sigma_{u'u'}$ and $\sigma_{v'v'}/\sigma_{u'u'}$. In the
critical zone $\sigma_{u'u'}$ and $\sigma_{yy}$ are indeed 
the same and the value $\sigma_{v'v'}$ is about $10$\% smaller.

\section{Conclusions}

We studied shear flows in a linear split-bottom cell by means of computer
simulations. The formation of wide shear zones was analyzed in the presence
and in the absence of gravity. In the former case pressure scales with
depth, in the latter case it is approximately constant. However, in both
cases the same type of wide shear zones emerge.

We showed that the widening of the shear zone in the bulk can be described
by one master curve which holds for various sizes and pressure
conditions. The shape of the widening function is a quarter of a circle and
not a power law as was suggested before.
We hope that this result will promote the development of the proper
continuum theory for quasi-static flows.

We analyzed the persistent growth of the critical zone and its effect on
the rheology. It influences the velocity field and the stresses, and
it causes a 
transient of the shear zone at the beginning of the shear test.

It was shown that the form of the stress tensor becomes
simpler with increasing shear strain $\gamma$. 
The region, where the stress tensor takes the reduced form
(\ref{eq-stress}), was also analyzed by Depken et al.\
\cite{Depken06b}. The authors found that this region can be best
characterized by the inertial parameter $I$ \cite{GDRMiDi04} which is
defined to be proportional to the shear rate $\dot\gamma$ and to the
inverse pressure. As it 
was pointed out in \cite{Depken06b} it is not clear how the emergence of
the inertial number can be reconciled with the rate independence of
quasi-static flows. It is the task of future work to clarify the question
what influence the parameters $\gamma$ and $I$ have on the stress
tensor in case of slow deformations.

\begin{acknowledgments}
We wish to thank J\'anos Kert\'esz for many suggestions and critical
remarks, and in particular for his hospitality extended to AR
during a research visit. We acknowledge partial support by
grant OTKA T049403, \"Oveges project GranKJ06 of
\includegraphics[height=1.5ex]{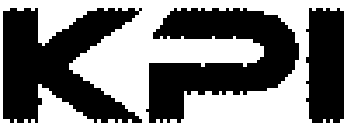} and 
\includegraphics[height=2.5ex]{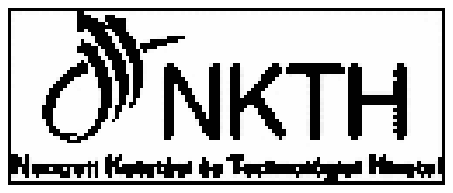} and the G.I.F. grant
No. I-795-166.10/2003. DEW thanks Francois Chevoir and Jean-Noel Roux for
their hospitality and many discussions on granular rheology during the fall
of 2005.
\end{acknowledgments}

\bibliography{granu}

\end{document}